\documentclass[reprint,superscriptaddress,amsmath,amssymb,amsfonts,amssymb,bm,aps,prl]{revtex4-2}

\usepackage{xcolor}
\usepackage{graphicx}
\usepackage{dcolumn}
\usepackage{bm}
\usepackage{hyperref}
\usepackage{physics}
\usepackage{times}
\usepackage{soul}
\hypersetup{
    colorlinks,
    linkcolor={blue!80!black},
    citecolor={blue!80!black},
    urlcolor={blue!80!black}
}
\usepackage[normalem]{ulem}
\definecolor{green}{RGB}{0,165,77}

\begin{document}

\title{Quantum survival in hybrid pursuit dynamics -- a lion chasing a lamb problem}

\author{P. Held}
\email{philip.held@uni-paderborn.de}
\affiliation{Paderborn University, Integrated Quantum Optics, Institute for Photonic Quantum Systems (PhoQS), Warburger Str. 100, 33098 Paderborn, Germany}
\author{\.{I}. Yal\c{c}{\i}nkaya}
\affiliation{Department of Physics, Faculty of Nuclear Sciences and Physical Engineering, Czech Technical University in Prague, B\v{r}ehov\'a 7, 115 19 Praha 1-Star\'e M\v{e}sto, Czech Republic}
\author{F. Pegoraro}
\affiliation{Paderborn University, Integrated Quantum Optics, Institute for Photonic Quantum Systems (PhoQS), Warburger Str. 100, 33098 Paderborn, Germany}
\author{J. Lammers}
\affiliation{Paderborn University, Integrated Quantum Optics, Institute for Photonic Quantum Systems (PhoQS), Warburger Str. 100, 33098 Paderborn, Germany}
\author{F. Schlue}
\affiliation{Paderborn University, Integrated Quantum Optics, Institute for Photonic Quantum Systems (PhoQS), Warburger Str. 100, 33098 Paderborn, Germany}
\author{K. Jir\'akov\'a}
\affiliation{Paderborn University, Integrated Quantum Optics, Institute for Photonic Quantum Systems (PhoQS), Warburger Str. 100, 33098 Paderborn, Germany}
\author{S. Barkhofen}
\affiliation{Paderborn University, Integrated Quantum Optics, Institute for Photonic Quantum Systems (PhoQS), Warburger Str. 100, 33098 Paderborn, Germany}
\author{B. Brecht}
\affiliation{Paderborn University, Integrated Quantum Optics, Institute for Photonic Quantum Systems (PhoQS), Warburger Str. 100, 33098 Paderborn, Germany}
\author{M. \v{S}tefa\v{n}\'ak}
\affiliation{Department of Physics, Faculty of Nuclear Sciences and Physical Engineering, Czech Technical University in Prague, B\v{r}ehov\'a 7, 115 19 Praha 1-Star\'e M\v{e}sto, Czech Republic}
\author{V. Poto\v{c}ek}
\affiliation{Department of Physics, Faculty of Nuclear Sciences and Physical Engineering, Czech Technical University in Prague, B\v{r}ehov\'a 7, 115 19 Praha 1-Star\'e M\v{e}sto, Czech Republic}
\author{A. G\'abris}
\affiliation{Department of Physics, Faculty of Nuclear Sciences and Physical Engineering, Czech Technical University in Prague, B\v{r}ehov\'a 7, 115 19 Praha 1-Star\'e M\v{e}sto, Czech Republic}
\affiliation{Institute for Solid State Physics and Optics, HUN-REN Wigner Research Centre for Physics, P.O. Box 49, H-1525 Budapest, Hungary}
\author{I. Jex}
\affiliation{Department of Physics, Faculty of Nuclear Sciences and Physical Engineering, Czech Technical University in Prague, B\v{r}ehov\'a 7, 115 19 Praha 1-Star\'e M\v{e}sto, Czech Republic}
\author{C. Silberhorn}
\affiliation{Paderborn University, Integrated Quantum Optics, Institute for Photonic Quantum Systems (PhoQS), Warburger Str. 100, 33098 Paderborn, Germany}

\date{\today}

\begin{abstract} 
Measurement and classical randomness usually degrade coherence; controlling where and when a subsystem is monitored can instead make them a tunable resource. We introduce a photonic time-multiplexed quantum walk whose fast reconfigurability enables precise position measurements, allowing us to study this interplay in a hybrid quantum-classical pursuit problem: A quantum walker (lamb) evolves on a line while a classical pursuer (lion) performs a lazy random walk. Each lion trajectory drives a distinct coherent, norm-reducing evolution of the lamb, with capture implemented by measurements at the lion's positions; averaging over the recorded trajectories then synthesizes a decoherent process. Counterintuitively, increasing the lion's mobility does not monotonically suppress survival: over a broad range of hopping probabilities, the lamb survives better than against an immobile predator. Directly observing this local enhancement, we demonstrate a manifestly quantum feature aided rather than hindered by classical stochastic control.
\end{abstract}

\maketitle

\paragraph*{Introduction---}
Coherence is a central resource in quantum information processing, enabling interference effects underlying quantum simulation and computational speedups over classical approaches~\cite{Streltsov2017,Aaronson2011}. Yet coherence is fragile: Unavoidable coupling to an environment degrades superpositions, progressively converting pure states into mixed states~\cite{Zurek2003}. While small systems can be well isolated, scaling remains challenging, as errors from multiple noise channels accumulate and degrade global correlations~\cite{Preskill2018,Schlosshauer2019,Cross2019}. Consequently, recent research has gone beyond error suppression to externally imposed randomness and measurement-driven dynamics, where controlled interplay between unitary evolution and the environment can produce phenomena such as phase transitions~\cite{Skinner2019,Li2018,Fisher2023} and anomalous survival dynamics in stochastic quantum processes~\cite{Stefanak2026recurrence}. Characterizing how repeated measurements and classical randomness shape the evolution and transport properties of realistic quantum systems is therefore a timely and significant open problem.

A natural setting to study competing coherent and incoherent processes is to subject a simple dynamical quantum system whose behavior is fully understood to an environment with an externally controlled random parameter. For example, a freely propagating particle in the presence of a single random scatterer -- or even an absorber which then is a possible quantum formulation of the pursuit problem.
The classical pursuit problem has a rich history: It originally modeled predator–prey (“lion–lamb”) chases, where geometry and motion strategies are crucial. A central question is whether a pursuer can always capture an evader in a bounded domain when both move at the same speed with perfect foresight~\cite{littlewood1986}. Optimal strategies guarantee capture while minimizing the evader’s maximum escape distance, though capture time can diverge~\cite{croft1964,flynn1973,flynn1974,flynn1974some,lewin1986,sgall2001}. This scenario inspired diffusive-capture models~\cite{krapivsky1996,redder1999} in which predators and prey perform random walks (RWs), where capture dynamics is determined by whether the relative walk is recurrent or transient, a well-known solved problem~\cite{Polya1921}.
Independent predators increase capture probability, inducing phase-transition-like behavior for three or more predators and finite-time capture in low dimensions~\cite{krapivsky1996,redder1999}.

Discrete-time quantum walks QWs~\cite{Aharonov1993,Meyer1996,Farhi1998}, the quantum counterparts of RWs, provide a natural framework for generalizing pursuit problems to the quantum domain.
They differ fundamentally from RWs and enable quantum advantage in tasks such as quantum Monte Carlo~\cite{Montanaro2015} and optimization~\cite{Marsh2020,Slate2021,Casares2022}, typically yielding quadratic speed-up over classical methods~\cite{Aaronson2003,Shenvi2003,Childs2004,Potocek2009,Melnikov2019,Ambainis2020,Apers2022,Oriekhov2024}.
Experimentally, photonic realizations have been demonstrated across diverse platforms, including bulk optics \cite{Bulk,Freq,UFTB,Fenwick}, structured light and orbital angular momentum modes \cite{Trans,OAM,esposito2022quantum}, fiber-loops \cite{Boutari, Bisianov, Dhinwa, Feis} as well as hybrid time-polarization architectures \cite{JCash,Pegoraro2023,CNOT}, and integrated interferometric circuits \cite{Sansoni2012,Crespi2013,CHIP}, enabling applications in quantum simulation and quantum information processing.

Studying pursuit problems with QWs can be viewed as a generalization of recurrence~\cite{Grunbaum2013}, with broader implications for theory and experiment. In the recurrence setting, the walker is monitored at the origin at each step and the walk is terminated upon detection, corresponding to a pursuit with a stationary predator. Such partial measurement induces conditional dynamics, with recurrence given by one minus the long-time survival probability. Recurrence of QWs on a line has been experimentally demonstrated using photonic time-multiplexing~\cite{Nitsche2018} and bulk optics~\cite{Chen2024}. Recent advances in electro-optical modulators (EOMs) and detector systems enable fast, reconfigurable monitoring of arbitrary positions during photonic time-multiplexed QWs, together with rapid data acquisition. Varying the monitored position between steps thus generalizes recurrence experiments to more complex pursuit dynamics.

Here, we experimentally implement a hybrid quantum–classical pursuit problem, revealing intricate open-system quantum dynamics.
The lamb performs a QW on a line while the lion executes a lazy RW, moving left/right with probability $\frac{p}{2}$ or staying put with probability $1-p$. The lamb is directly represented by a wave packet spreading in a time-multiplexed QW setup, while the lion's position is dynamically encoded using programmable EOMs such that the capture is implemented by deterministic out-coupling of time-bins corresponding to the lion's position.

Intuitively, a spreading lion should reduce the lamb's survival compared to an immobile lion ($p = 0$) due to covering more space, aligning better with the lamb's positions, and the classical randomness competing with the QW's coherence.
Surprisingly, as $p$ increases from 0 to 1, we observe that the lamb's survival probability does not decrease monotonically; for a broad range of $p$, it is higher than with an immobile lion, due to an intricate interplay between the lamb's coherent, ballistic spreading, the lion's strict localization, and the statistical distribution of its trajectories.

\paragraph*{Theory---}

The lamb's QW in the absence of any absorption is governed by $\ket{\Psi(t+1)}=\hat U\ket{\Psi(t)}$, where the unitary $\hat U$ acts on a bipartite coin-position Hilbert space $\mathcal{H}_c \otimes \mathcal{H}_p$  spanned by $\ket{c,x}$ with $c\in\{-1,+1\}$ and $x\in \mathbb{Z}$. The operator $\hat U~=~\hat S(\hat C\otimes \hat I_p)$ defines one step of the walk, consisting of a coin operator $\hat C$ and a shift operator $\hat S$. $\hat C$ acts only on the coin space, and we choose it as the Hadamard transform.
$\hat{S}=\sum_{x,c} \ketbra{c}{c}\otimes\ketbra{x+c}{x}$ acts on the total Hilbert space and moves the walker to the left (right) for the coin state $\ket{-1}$ ($\ket{+1}$). After $t$ steps, the final state takes the general form $\ket{\Psi(t)}= \hat U^t\ket{\Psi_0}=\sum_{c,x}\psi_{c,x}(t)\ket{c,x}$, and the probability of finding the walker in the state $\ket{c,x}$ becomes $P(x,c,t)=|\psi_{c,x}(t)|^2$.

\begin{figure}[t]
	\centering
	\includegraphics[width=1\linewidth]{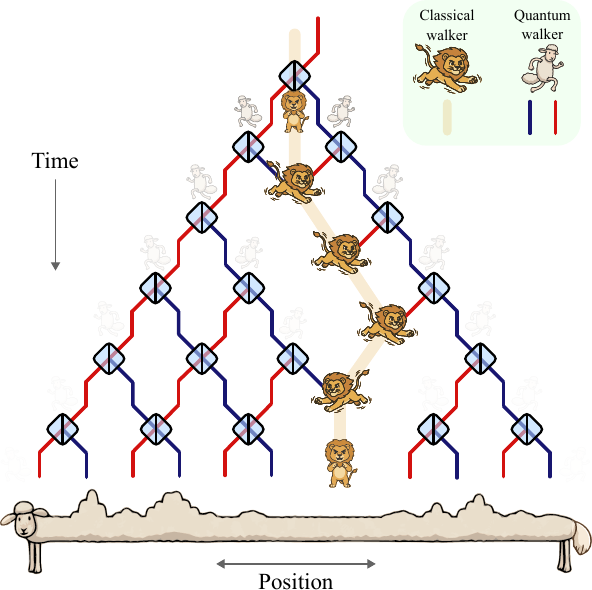}
	\caption{Pictorial illustration of the hybrid quantum-classical pursuit problem over six steps using a quantum analogue of a Galton board. The lamb (a photon), initialized in the state $\ket{-1,0}$, propagates through an array of 50/50 beam splitters. The light-brown track indicates one possible trajectory of the lion, whose successive projections of the lamb's wavefunction give rise to the probability distribution reconstructed from many repetitions.}
	\label{fig:concept}
\end{figure}

Both the quantum lamb and the classical lion start at the origin, with no absorption initially (see Fig.~\ref{fig:concept}). Each step consists of (i) the lamb's QW evolution via $\hat U$, (ii) the lion's stochastic diffusion according to a lazy RW, and (iii) projection of the lamb's state onto the subspace excluding the lion’s position, making the overall dynamics non-unitary. The lion stays in place with probability $p_0$ or moves left/right with probabilities $p_{\pm 1}$. We consider a symmetric case
\begin{equation}
p_0 = 1-p,\quad  p_{\pm 1} = \frac{p}{2}, \quad 0\leq p\leq 1.
\label{eq:jumpProbs}
\end{equation}
Changing $p$ allows for interpolation between a static lion ($p=0$) and a simple balanced RW ($p=1$). For the case of a static lion, the pursuit problem is equivalent to the recurrence of a QW \cite{Grunbaum2013,bourgain_quantum_2014}.

We examine the quantum lamb's probability of not being absorbed by the lion during evolution, i.e., the \textit{survival probability} $S$. Consider a given realization of the lion's RW $r$ with trajectory $x_r: \mathbb{N} \to \mathbb{Z}$. Erasure of the lamb's amplitude in the lion's position $x_r(i)$ is given by a projection 
\begin{equation}
\hat \pi_{x_r(i)} = \hat{I}_c\otimes(\hat I_p - \ketbra{x_r(i)}{x_r(i)}).   \end{equation}
For a lion's trajectory $x_r$, the lamb's state after $t$ steps is given by
\begin{equation}
\label{psi:r}
\ket{\Psi_r(t)} = \hat{\pi}_{x_r(t)}\hat{U} \hat{\pi}_{x_r(t-1)}\hat{U} \cdots \hat{\pi}_{x_r(1)}\hat{U} \ket{\Psi_0},
\end{equation}
where $\ket{\Psi_0} = \ket{\psi_c}\otimes\ket{0}$ with initial coin state $\ket{\psi_c}$.
The expression $|\psi_{c,x}(t)|^2 = |\braket{c,x}{\Psi_r(t)}|^2$ retains the meaning of probability of finding the lamb. Thus, the sum over all $x$ and $c$, $S_r(t) = \lVert \Psi_r(t)\rVert^2$, is not unity, but the survival probability until time $t$ for the given realization $r$. Note that $S_r(t)$ also depends on the choice of $\ket{\psi_c}$, which will be left implicit in the notation.
Averaging over all $n(t)$ possible lion's trajectories yields the lamb's survival probability
\begin{equation}
S_t(p) = \sum_{r=1}^{n(t)} P_r S_r(t),
\label{eq:esPrAv}
\end{equation}
where $P_r$ is the probability of trajectory $x_r$. 
This is completely determined by the jump probabilities \eqref{eq:jumpProbs}.
Note that, unlike $S_r(t)$, $S_t(p)$ is \emph{not} a function of the initial coin state $\ket{\psi_c}$. The justification is provided in the Supplementary material.

\begin{figure}[b]
	\centering
    \includegraphics[width=0.98\linewidth]{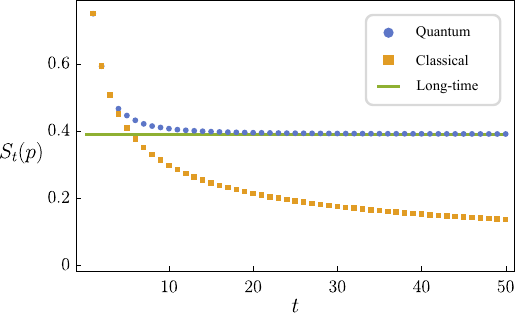}
	\caption{Comparison of the survival probability for the quantum-classical pursuit problem and the fully classical case for $p=1/2$. In the quantum-classical one, the long-time behavior (green line) is reached within 15 to 20 steps sufficiently well, making particularly suited for experiential investigations with a photonic QW.}
	\label{fig:qxc}
\end{figure}

Figure \ref{fig:qxc} displays the course of the quantum lamb's survival probability (blue dots) in the first 50 steps for $p=1/2$. For comparison, we show the survival probability for a model in which both the lamb and the lion are classical (orange squares). One can clearly see that the two cases feature different behavior.
In the hybrid pursuit problem, $S_t(p)$ saturates at a nonzero value, whereas in the classical case it vanishes as $t^{-1/2}$.
For more details on the long-time limit of $S_t(p)$, we refer to the Supplementary material.

\paragraph*{Experiment---}

For an experimental investigation of $S_t(p)$, we propagate the QW over $20$ steps.
This balances the proximity to the long-time regime and the feasibility within existing photonic implementations.
To describe the lion's action, we introduce auxiliary modes (AMs).
Coupling a photon from the QW into an AM is equivalent to the lion catching the lamb.
The corresponding part of the lamb's wavefunction in the AM is erased from the QW evolution \cite{Nitsche2018}.

\begin{figure}[t]
	\centering
	\includegraphics[width=1\linewidth]{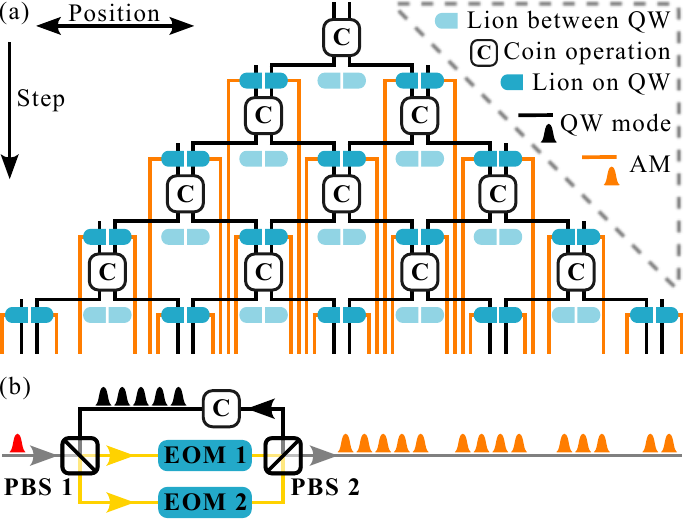}
	\caption{(a) Network for implementing the lion’s action by coupling each QW mode in each step to an auxiliary mode (AM). With increasing step number, the number of QW modes grows linearly whereas the number of AMs grows quadratically. (b) Schematics of the time-multiplexed QW setup. Via two fiber loops (yellow) and a feedback path (black) the lamb's evolution is implemented, while the EOMs are used to implement the lion's action. By separating QW modes and AMs, an excellent scaling behavior is achieved.}
	\label{fig:experiment}
\end{figure}

Figure \ref{fig:experiment}(a) schematically depicts the network required for implementing a quantum-classical hybrid system.
We show the mode structure of a pure QW in black with the coin operation acting locally on each occupied position and the successive position shift.
The light and dark teal elements depict the lion. 
Since the lion must not move in every step, it can end up on a position that is not occupied by the QW (light teal), where it cannot catch the lamb. 
If the lion ends up in an occupied position (dark teal), it redirects the lamb to an AM (orange lines). 
The number of necessary AMs scales quadratically with the step number in this sketch, resulting for $t=20$ in 380 AMs in addition to the 40 QW modes and with it the 380 corresponding couplings that must be set individually.
For a detailed analysis of the necessary number of AMs and the requirements on system reconfigurability, we refer to the End matter.

Such large-scale evolutions \cite{UFTB, Boutari, Bisianov} as well as the required reconfigurability \cite{CHIP, Dhinwa, Yu} have only been shown individually to date. 
Our photonic time-multiplexed QW \cite{JCash} addresses this challenge by naturally combining a large number of modes with fast reconfigurability.
It maps each output mode of the network from Fig. \ref{fig:experiment}(a) onto a well-defined arrival time bin, hence carries many individual modes in the same spatial channel.
In Fig. \ref{fig:experiment}(b), we present a schematic of the setup.
After coupling the light into the setup at polarizing beam splitter (PBS) 1 the QW modes evolve inside an unbalanced Mach-Zehnder interferometer (yellow), whose output loops back to its input (black).
Each roundtrip with duration $\tau$ corresponds to a single QW step.
Coupling to AMs is implemented by EOMs 1 and 2, which direct the light to the output (gray) at PBS 2.
Thus, the AMs are separated physically from the QW evolution, preventing back action.

We realize the quantum lamb with a coherent light pulse. 
This is justified because evaluating $S_r(t)$ requires only intensity measurements, i.e., first-order correlations, which are the same for coherent states and true single photons \cite{Paul}.
We encode the coin state in polarization, with $\ket{\psi_c}=\ket{-1}$ being vertically polarized.
The Hadamard coin operation in the QW evolution (c.f. Fig. \ref{fig:experiment}(b)) is realized with a half-wave plate.
Then, the step is realized with two fibers of different lengths that form the two arms of the unbalanced Mach-Zehnder.
They introduce a relative time difference $\Delta\tau\ll\tau$ that we call position separation. 
This is the timing between successive pulses within one step.

The lion's action corresponds to setting the EOMs to couple all light at the lion's positions out of the QW evolution into the corresponding AMs. 
Our EOMs feature switching times of $20\,\mathrm{ns}$, sufficient for addressing individual positions. 
Besides implementing the lion, we also use the two EOMs to deterministically couple the walker into and out of the QW evolution.
The setup's total reconfiguration time, that is, the time for switching from trajectory $x_r$ to $x_{r+1}$, is two seconds, allowing us to probe many trajectories in a short time. 

We detect the walker's output probability distribution after 20 steps with superconducting nanowire single-photon detectors (SNSPDs), where click arrival times map to QW positions.
To increase acquisition rates, we use a 1-to-8 binary fiber tree to multiplex the signal onto eight SNSPDs. 
To avoid detector saturation, we set the energy of the coherent pulse such that, on overage, not more than 0.01 photons hit each detector in every time bin. 
Finally, to also resolve the coin degree of freedom, we separate H and V polarizations at the setup's output and use one detection unit per polarization. 

We scan $p$ from 0 to 1 with a resolution of 0.1.
For each $p$, we sample 200 random trajectories.
For each trajectory, we integrate for one second, which, given an overall clock rate of $19\,\mathrm{kHz}$, corresponds to 19,000 individual QW realizations. 
We record up to 5,000 individual clicks per trajectory, from which we obtain $S_r(t=20)$ as the ratio of detected clicks and clicks of a reference QW without lion.
Finally, we obtain $S_{t=20}(p)$ as the arithmetic mean of all $S_r(t=20)$.
Note that the statistical weights of the individual trajectories have already been accounted for by the random sampling from their distribution.

\paragraph*{Results---}

\begin{figure}[t]
	\centering
	\includegraphics[width=1\linewidth]{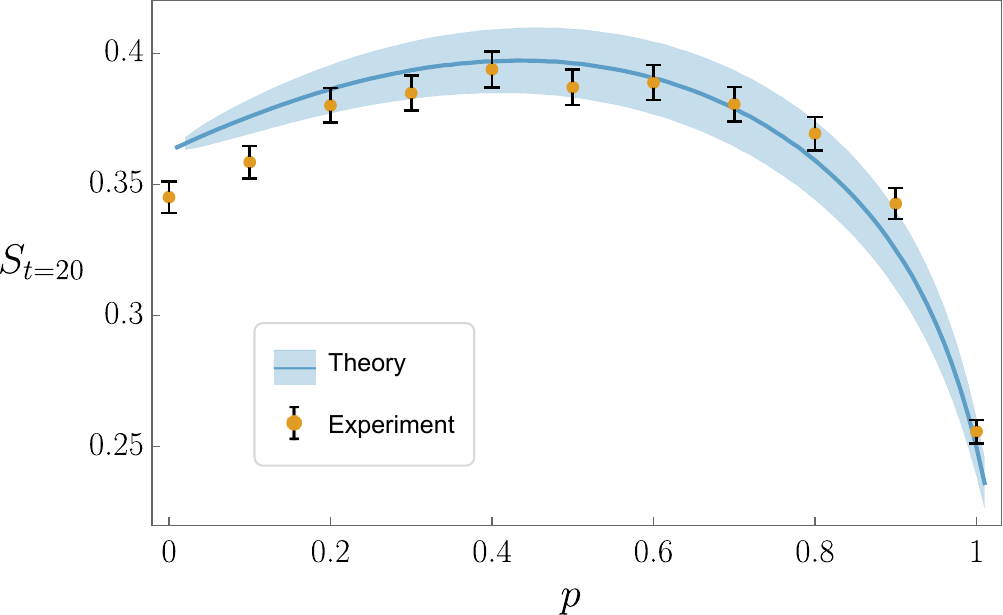}
	\caption{Measured survival probability at step 20 compared to theory for varying $p$. Counterintuitively, the survival probability of lamb initially increases despite the increasing randomness of the lion's movement, demonstrating that mixing classical stochasticity and quantum evolution can enhance survival through modified interference, resembling the non-monotonic recurrence behavior of stochastic walks~\cite{Stefanak2026recurrence}. The blue area indicates one sigma uncertainty due to finite sampling of trajectories.}
	\label{fig:results}
\end{figure}

Figure \ref{fig:results} reports the experimental (orange markers) and analytical (blue line) survival probability as function of the hopping probability.
The blue area around the theoretical trend indicates one-sigma uncertainty (below 4.8\%) due to sampling a limited set of 200 trajectories.
The uncertainties of the experimental data contain statistical uncertainties (below 0.25\%), detector saturation effects (below 0.5\%), and detector efficiency fluctuations due to polarization rotations in the fibers leading to the detectors (below 1.7\%).
Details on the derivation of uncertainties are given in the End matter.
Experiment and theory are in good agreement, allowing us to investigate effects of open quantum systems via the hybrid pursuit problem.
The main mismatch between theory and experiment is visible for $p=0$ and $p=0.1$.
Here, the required EOM switching patterns are very similar for different trajectories due to the lion's low hopping probability.
We could observe that this causes a non-ideal implementation of the applied switching patterns, most probably due to acoustic resonances, a known phenomenon \cite{Thomaschewski} impacting the performance of Pockels cells.
From $p=0.2$ on, the patterns feature sufficient differences for a reliable implementation.

For $p=0$, the lion stays at $x=0$ and, following the QW dynamics, can catch the lamb only in even steps.
In contrast, for $p=1$, the lion moves in every step, just like the lamb, and has a likelihood to chase down the lamb in every step. 
Hence, we expect $S_{t=20}(p=0) > S_{t=20}(p=1)$, which can be seen both in theory and experiment.

Counterintuitively, $S_{t=20}(p)$ does not decrease monotonously from $p=0$ to $p=1$.
We observe an initial increase until $p\approx0.45$, then a similar decrease until $p\approx0.8$ , followed by a rapid decrease until $p=1$.
Due to the intricate interplay of classical and quantum evolutions, the interference in the QW is modified and increases the survival probability, even if a purely classical overlap argument would suggest the opposite.
Note that we investigated theoretically the recurrence probability in discrete-time quantum stochastic walks in \cite{Stefanak2026recurrence}.
We showed that it does not monotonously increase from a fully quantum to fully classical walk, but in fact features an initial decrease.
Observing now a very similar effect in a related, but fundamentally different, physical system supports a broader validity of this behavior and hints at the possibility of future generalizations.

The enabling feature for investigating the hybrid pursuit problem was the availability of AMs due to direct out-coupling from individual positions of the evolution and the quick and arbitrary reconfiguration of our EOMs.
In Tab. \ref{tab:comparison} we compare existing photonic QW implementations in light of their suitability for this work.
As it stands, only our time-multiplexed platform provides the required number of QW modes and AMs, even for drastically reduced requirements in other platforms (see End matter).
Furthermore, not all platforms satisfy the hard requirement of individual reconfigurability necessary to sample many different trajectories. 
Of those that do, only one (\cite{Freq}) features reconfiguration times similar to our platform and therefore can support reasonable measurement times. 
This cements our platform as promising candidate to advance this research.

\begin{table}[t]
    \centering
    \caption{Comparison of photonic QW realizations regarding their suitability for studying pursuit problems: transverse momentum (Trans), orbital angular momentum (OAM), frequency, ultrafast time-bins (UFTB), bulk optics, integrated chips, and this work: time-multiplexing.
    A suitable realization needs to provide enough evolution modes (EMs) and AMs and individual reconfigurability, i.e., position-resolved, of the couplings between evolution and AMs.
    In case of future scaling of mode sizes, the required total run time varies strongly due to different reconfiguration mechanisms (frequency and time-multiplexing: EOMs, integrated chips: heaters, and bulk: manual operation).}
    \label{tab:comparison}
    \begin{tabular}{|c|c|c|c|c|}
        \hline
        Realization &  Required & Available & Indiv. reconf. & Total \\
         & EM (AM)  & EM (AM) & (time) &  runtime \\
        \hline
        \hline
        Trans \cite{Trans} & 20(19) & \textcolor{red}{5(15)} & \textcolor{red}{No} & n.a. \\
        \hline
        OAM \cite{OAM} & 20(19) & \textcolor{red}{7(0)} & \textcolor{red}{No} & n.a. \\
        \hline
        Frequency \cite{Freq} & 20(19) & \textcolor{red}{20(13)} & \textcolor{green}{Yes} (2 sec) & Hours \\
        \hline
        UFTB \cite{UFTB} & 40(380) & \textcolor{red}{100(0)} & \textcolor{red}{No}  & n.a. \\
        \hline
        Bulk \cite{Bulk} & 40(380) & \textcolor{red}{12(42)} & \textcolor{green}{Yes} (10 min) & Weeks \\
        \hline
        Chip \cite{CHIP} & 40(38) & \textcolor{red}{20(0)} & \textcolor{green}{Yes} (30 sec) & Days \\
        \hline
        \hline
        This work & 40(380) & \textcolor{green}{64(922)} & \textcolor{green}{Yes} (2 sec) & Hours \\
        \hline
    \end{tabular}
\end{table}

\paragraph*{Conclusion---}
By implementing the hybrid lion and lamb pursuit problem, we demonstrated how flexible and scalable QWs can be used to study monitored quantum dynamics, i.e., dynamics in which a classical stochastic evolution (lion) dictates where a quantum evolution (lamb) is measured, and whose incoherent character emerges only upon averaging over the lion's trajectories. We find the counterintuitive behavior that the quantum system's survival probability is not a monotonically decreasing function of the lion's mobility. In fact, we observe it increased for a wide range of hopping probabilities. The good agreement of measurements with theoretical predictions highlights the suitability of our time-multiplexing platform as quantum simulator for the intricate interplay between coherent quantum evolution, measurement back-action, and classical randomness, which leads to unexpected effects and results. Thanks to our system size and fast reconfiguration speed, we were able to measure more than 4,000 individual configurations in a few hours with a statistical error below 0.3\%.

For further investigation, the experimental setup can readily be adapted to study, on the one hand, variations of hybrid pursuit problems such as using multiple lions, introducing eating probabilities, or applying dynamic coin operations \cite{CNOT} and, on the other hand, new problems requiring experimental capabilities such as coupling to a thermal bath to introduce genuine environment-induced decoherence, storing and coupling back out-coupled light, or using a variety of quantum input states \cite{Nitsche2020, Pegoraro2023}. In particular, the extensive body of work on classical multi-walker systems naturally suggests the existence of rich quantum counterparts, where interference and correlations may give rise to qualitatively new and counterintuitive collective phenomena. We therefore expect the present framework to provide a promising platform for exploring multipartite hybrid classical/quantum dynamics and uncovering genuinely quantum transport effects beyond the studied two-particle regime.

\paragraph*{Acknowledgments---}
The authors thank Michael Stefszky for valuable feedback on the manuscript.
The authors acknowledge financial support by the European Commission through the Horizon Europe project EPIQUE (Grant No. 101135288).
P.H., F.P., J.L., F.S., K.J., B.B, and C.S. received funding from the German Federal Ministry of Research, Technology and Space (BMFTR) within the PhoQuant project (Grant No. 13N16103).
F.S. is part of the Max Planck School of Photonics supported by the Dieter Schwarz Foundation, the German Federal Ministry of Research, Technology and Space (BMFTR), and the Max Planck Society.
A. G. and I. J. have been supported by the Grant Agency of the Czech Republic GAČR under Grant No. 26-23973S.
I. J., \.{I}. Y., M. Š., V. P. have been supported by the Grant Agency of the Czech Republic GAČR under Grant No. 23-07169S.
The funders played no role in study design, analysis and interpretation of data, or the writing of this manuscript. 

\paragraph*{Data availability---}
The measured click data and the codes for reconstructing the survival probability and the simulation are openly available \cite{code}.

\appendix

\section{End matter}

\paragraph*{Required mode size of experimental systems---}

First, the QW evolution needs to be implemented.
A QW on a line with a localized input state spreads over $2t+1$ positions after $t$ steps.
The required mode size is twice the number of positions if the coin state is a physical quantity and not implemented only in the coin operation.
Due to the structure of a QW, namely unoccupied even and odd positions in odd and even steps, respectively, and unoccupied coin modes in the outermost positions in the last step, the problem can be reduced to $2t$ modes and thus be implemented by a 40×40 unitary evolution for 20 steps.

In addition to this evolution, auxiliary modes for the lion are required.
In the case of a static evolution, due to the different trajectories of the lion, coupling to an individual auxiliary mode at each position in each step needs to be possible.
Thus, the implementation of the QW evolution itself stays unchanged and only the coupling to the auxiliary modes is altered.
For the last step no auxiliary modes are required, as the walker is measured at this point.
Thus, the demand for auxiliary modes scales as $\sum_{t'=1}^{t'=t-1}(2t) = t^2-t$.
For our QW with 20 steps, this corresponds to 380 auxiliary modes.

To overcome the undesirable $t^2$ scaling one can harness the flexibility of fully reconfigurable networks.
Allowing for reconfiguring the network in a way that not only the coupling from the QW modes to the auxiliary modes is altered but also that the modes of the evolution are reshuffled for the different trajectory implementations, the demand for auxiliary modes can be reduced to two per step.
Thus, the demand scales as $2t-2$, resulting in 38 auxiliary modes for a 20 step QW.
Note that in case of implementations with the coin state being only realized in the coin operation, the required number of evolution as well as auxiliary modes is halved.

\paragraph*{Data reconstruction and uncertainties---}

The survival probability for a lion's trajectory $x_r$ is calculated from the raw clicks $C$ summed over the 8 detectors per coin state as
\begin{equation}
S_r(t=20) = \frac{C_{r,\mathrm{H}} + C_{r,\mathrm{V}}}{C_\mathrm{ref,H} + C_\mathrm{ref,V}},
\label{eq:SfromC}
\end{equation}
with the indices denoting the trajectory ($r$) or reference QW without a lion ($\mathrm{ref}$) and the coin state ($\mathrm{H}$ or $\mathrm{V}$).
Averaging over all $R$ trajectories for a given $p$ yields 
\begin{equation}
S_{t=20} = \frac{\sum\limits_{r=1}^{R}S_r(t=20)}{R}.
\end{equation}
Note that the $p$-dependence of $S_t$ is reflected in the probability distribution of the trajectories, so we drop the explicit notation.

Substituting $C_r = C_{r,\mathrm{H}} + C_{r,\mathrm{V}}$ and $C_\mathrm{ref} = C_\mathrm{ref,H} + C_\mathrm{ref,V}$, the statistical uncertainties of the clicks assuming Poissonian counting statistics are propagated via Gaussian error propagation to
\begin{equation}
\Delta_{S_r(t=20),\mathrm{stat}} = \sqrt{\left (\frac{\sqrt{C_r}}{C_\mathrm{ref} }\right )^2 + \left (\frac{C_r\sqrt{C_\mathrm{ref}}}{C_\mathrm{ref}^2}\right )^2}.
\end{equation}
This yields
\begin{equation}
\Delta_{S_{t=20},\mathrm{stat}} = \frac{\sqrt{\sum\limits_{r=1}^{R}\Delta_{S_r(t=20),\mathrm{stat}}^2}}{R}.
\end{equation}

The error due to detector saturation arises from the Poissonian photon number distribution of the coherent light and the non-photon-number-resolving nature of the SNSPDs used.
Higher photon number contributions are detected as a single click, leading to an underestimation of counts.
The probability for a $n$ photon event at a mean photon number of $\bar n$ is given by
\begin{equation}
P(n) = \frac{\bar n ^n}{n!}e^{-\bar n},
\end{equation}
with $\sum_{n=0}^\infty P(n)= 1$.
Thus, with the number of trials $N$ the expectation value for the recorded clicks reads as
\begin{equation}
\mathbb{E}[C_\mathrm{rec}] = N\sum\limits_{n=1}^\infty P(n)\cdot 1 = N(1-e^{-\bar n})
\end{equation}
and the expectation value for the non-recorded clicks as
\begin{equation}
\mathbb{E}[C_\mathrm{non-rec}] = N\sum\limits_{n=1}^\infty P(n)\cdot (n-1) = N[\bar n - (1-e^{-\bar n})].
\end{equation}
From this, we calculate the relative error in detected counts as
\begin{equation}
\Delta_\mathrm{count} = \frac{\mathbb{E}[C_\mathrm{non-rec}]}{\mathbb{E}[C_\mathrm{rec}]} = \frac{\bar n}{1-e^{-\bar n}} - 1.
\end{equation}

This error is a systematic bias for each recorded mode in each measurement.
However, as the mean photon number in each mode varies due to changing probability distributions of the lamb, we derive an upper bound for this error.
Thus, we analyze the counts at the single-detector, single-QW-position and individual-coin level.
We extract as maximum mean photon number 0.0084.
As the bias reduces the recorded counts, its appearance in the numerator and denominator in Eq. \ref{eq:SfromC} partially cancels between the two.
The amount of this effect depends on the individual trajectory and the following change in probability distributions of the lamb.
Thus, to obtain the upper bound for this error, we assume only an error on the reference QWs and no error on the QWs with lion, erasing the counteraction.
Gaussian error propagation for a single trajectory yields
\begin{align}
\nonumber \Delta_{S_r(t=20),\mathrm{count}} &= \sqrt{\left (\frac{C_r}{C_\mathrm{ref}^2}C_\mathrm{ref}\Delta_\mathrm{count}\right )^2} \\
&= S_r(t=20) \Delta_\mathrm{count},
\end{align}
where we used that the absolute error follows from the relative as $C_\mathrm{ref}\Delta_\mathrm{count}$.
Note that assuming only an error on the QWs with lion and no error on the reference QWs would yield the same result.
As the errors for the individual trajectories are caused by a systematic bias, they are treated as fully correlated systematic and therefore propagated linearly
\begin{equation}
\Delta_{S_{t=20},\mathrm{count}} = S_{t=20}\Delta_\mathrm{count}.
\end{equation}

Due to polarization rotation in the fibers to the detectors and slight differences in absolute detection efficiencies, a detection efficiency imbalance between the two coin states can occur.
This can be accounted for by introducing in Eq. \ref{eq:SfromC} the ratio of efficiency for H and V detection $q$, generalizing to
\begin{equation}
S_{r,\mathrm{bal}}(t=20) = \frac{q C_{r,\mathrm{H}} + C_{r,\mathrm{V}}}{q C_\mathrm{ref,H} + C_\mathrm{ref,V}}.
\end{equation}

For error propagation the deviation $\Delta_q$ from $q=1$ is considered,
\begin{widetext}
\begin{equation}
  \begin{aligned}
    \Delta_{S_r(t=20),\mathrm{bal}} &= \sqrt{\left( \frac{\partial S_r(t=20)}{\partial q} \right)^2_{q=1} (\Delta_q)^2} = \left| \frac{\partial S_r(t=20)}{\partial q} \right| \Delta_q \\
    &= \left| \frac{C_{r,\mathrm{H}}}{q C_\mathrm{ref,H} + C_\mathrm{ref,V}} - \frac{q C_{r,\mathrm{H}} + C_{r,\mathrm{V}}}{(q C_\mathrm{ref,H} + C_\mathrm{ref,V})^2} C_\mathrm{ref,H} \right|_{q=1} \Delta_q \\
    &= \left| \frac{C_{r,\mathrm{H}}}{C_\mathrm{ref,H} + C_\mathrm{ref,V}} - \frac{C_{r,\mathrm{H}} + C_{r,\mathrm{V}}}{(C_\mathrm{ref,H} + C_\mathrm{ref,V})^2} C_\mathrm{ref,H} \right| \Delta_q \\
    &= \left| \frac{C_{r,\mathrm{H}}}{C_\mathrm{ref}} - \frac{C_r}{C_\mathrm{ref}^2} C_\mathrm{ref,H} \right| \Delta_q, \\
  \end{aligned}
\end{equation}
\end{widetext}
which can be written as
\begin{equation}
  \begin{aligned}
    \Delta_{S_r(t=20),\mathrm{bal}} &= \frac{C_r}{C_\mathrm{ref}} \left| \frac{C_{r,\mathrm{H}}}{C_r} - \frac{C_\mathrm{ref,H}}{C_\mathrm{ref}} \right| \Delta_q \\
    &= S_{r}(t=20) \left| \frac{C_{r,\mathrm{H}}}{C_r} - \frac{C_\mathrm{ref,H}}{C_\mathrm{ref}} \right| \Delta_q.
  \end{aligned}
\end{equation}
Both terms, $\frac{C_{r,\mathrm{H}}}{C_r}$ and $\frac{C_\mathrm{ref,H}}{C_\mathrm{ref}}$, can vary between 0 and 1 depending on the input state and $\frac{C_{r,\mathrm{H}}}{C_r}$ additionally depending on the chosen trajectory.
Thus, we calculate the average error with $\int_0^1 dx \int_0^1 dy\,|x-y| = \frac{1}{3}$ as 
\begin{equation}
\Delta_{S_r(t=20),\mathrm{bal}} = S_{r}(t=20)\frac{\Delta_q}{3}.
\end{equation}
In the worst case, if $\Delta_q$ does not vary during measurement, this error is a systematic bias, thus we propagate them linearly
\begin{equation}
\Delta_{S_{t=20},\mathrm{bal}} = S_{t=20}\frac{\Delta_q}{3}.
\end{equation}
From the drifts observed experimentally we estimate $\Delta_q = 0.05$ $(1\sigma)$.

The uncertainty due to sampling from a limited set of trajectories is calculated as the standard error
\begin{equation}
\Delta_{S_{t=20},\mathrm{samp}} = \frac{\sigma_S}{\sqrt{R}},
\end{equation}
where $\sigma_S$ is the standard deviation of $S_r(t=20)$ in the limit of infinite trajectories.
We simulate $10^7$ trajectories for each $p$ to approach the long-time regime and obtain $\sigma_S$.

The calculated errors are displayed in Tab. \ref{tab:error}.
It shows that from the experimental errors the one due to the detection imbalance is the dominant contribution, as the errors from statistics and saturation effects of the detectors are effectively minimized by the measurement design.
These three experimental errors are added in quadrature to obtain the error plotted in Fig. \ref{fig:results}.
However, even the combination of all three experimental errors is below half of the uncertainty due to the sampling statistics.

\begin{table}[t]
    \centering
    \caption{Comparison of the error contributions.}
    \label{tab:error}
    \begin{tabular}{|c||c|c|c||c|}
        \hline
        $p$ &  $\Delta_{S_{20},\text{stat}}$ & $\Delta_{S_{20},\text{count}}$ & $\Delta_{S_{20},\text{bal}}$ & $\Delta_{S_{20},\text{samp}}$\\
        \hline
        \hline
        0 & 0.0007 & 0.0015 & 0.0058 & 0 \\
        \hline
        0.1 & 0.0007 & 0.0016 & 0.0060 & 0.0070 \\
        \hline
        0.2 & 0.0008 & 0.0016 & 0.0064 & 0.0094 \\
        \hline
        0.3 & 0.0008 & 0.0017 & 0.0065 & 0.0110\\
        \hline
        0.4 & 0.0008 & 0.0017 & 0.0066 & 0.0122\\
        \hline
        0.5 & 0.0008 & 0.0017 & 0.0065 & 0.0132 \\
        \hline
	    0.6 & 0.0008 & 0.0017 & 0.0065 & 0.0139 \\
        \hline
        0.7 & 0.0008 & 0.0017 & 0.0064 & 0.0146 \\
        \hline
        0.8 & 0.0080 & 0.0016 & 0.0062 & 0.0150 \\
        \hline
        0.9 & 0.0008 & 0.0015 & 0.0058 & 0.0148 \\
        \hline
        1 & 0.0007 & 0.0011 & 0.0043 & 0.0096 \\
        \hline
        \hline
        $\Delta /  S_{20}$ & $<0.25\%$ & $<0.5\%$ & $<1.7\%$ & $<4.8\%$ \\
        \hline
    \end{tabular}
\end{table}

\bibliography{bibliography}

\end{document}